\documentclass[prb,twocolumn,showpacs,amsmath,amssymb,
               letterpaper,citeautoscript]{revtex4}

\usepackage[dvips]{graphicx}
\usepackage{dcolumn}
\usepackage{bm}
\usepackage{times}

\begin{document}


\title{
Models of Electrodes and Contacts in
Molecular Electronics }

\author{ San-Huang Ke,$^{1,2}$ Harold U. Baranger,$^{2}$ and Weitao Yang$^{1}$}

\affiliation{
     $^{\rm 1}$Department of Chemistry, Duke University, Durham, NC 27708-0354 \\
     $^{\rm 2}$Department of Physics, Duke University, Durham, NC 27708-0305
}

\date{5 June 2005; \textbf{J. Chem. Phys. 123, 114701 (2005)}; DOI: 10.1063/1.1993558}

\begin{abstract}
Bridging the difference in atomic structure between experiments and theoretical calculations and exploring quantum confinement effects in thin electrodes (leads) are both important issues in molecular electronics. To address these issues, we report here, by using Au-benzenedithiol-Au as a model system, systematic investigations of different models for the leads and the lead-molecule contacts: leads with different cross-sections, leads consisting of infinite surfaces, and surface leads with a local nanowire or atomic chain of different lengths. The method adopted is a non-equilibrium Green function approach combined with density functional theory calculations for the electronic structure and transport, in which the leads and molecule are treated on the same footing. It is shown that leads with a small cross-section will lead to large oscillations in the transmission function, $T(E)$, which depend significantly on the lead structure (orientation) because of quantum waveguide effects. This oscillation slowly decays as the lead width increases, with the average approaching the limit given by infinite surface leads. Local nanowire structures around the contacts induce moderate fluctuations in $T(E)$, while a Au atomic chain (including a single Au apex atom) at each contact leads to a significant conductance resonance.
\end{abstract}

\maketitle
\section{Introduction}

To have precise control over the atomic structure of individual molecular
devices, which consist of, at least, a molecule and two electrodes used as
leads of electronic current, is one of the major challenges in molecular
electronics. In recent experiments on electron transport through single
molecule devices, a break junction is a commonly used technique for building
a lead-molecule-lead (LML) system. A break junction can be constructed either
through electromigration \cite{morpurgo1,park,morpurgo2,fuhrer} or through
the mechanically controllable break junction (MCB) technique \cite{bj1,bj2,bj3,h2}.
In the MCB technique a metal wire is elongated and broken by the bending of the
substrate, and therefore the resulting break gap can be controlled by adjusting
the bending, providing a flexibility for controlling the device structure. In
all these break junction experiments, the detailed atomic structure of the
molecule-lead contacts is not available and so neither is its influence on the
transport properties of the device. However, some information about the main
features of the contact atomic structure has been revealed in MCB experiments.
It has been shown that well before a metal wire breaks a very thin bridge
region is formed which contributes only several $G_0$ (=$2e^2/h$, conductance
quantum) of conductance to the wire \cite{h2,wire}. This means that in a real
MCB LML system the molecule is usually connected to a very thin nanowire which
is then connected to the extended part of the metal lead.

Other experimental approaches for constructing well defined
lead/single-molecule/lead systems involve the use of chemical self-assembly of
molecules on surfaces and/or direct atomic manipulation using scanning
tunneling microscopy (STM) or atomic-force microscopy (AFM) (surface-STM/AFM
technique) \cite{Datta972530,Cui01571,Xiao04267}. In these experiments one of
the leads is an infinite large surface (the substrate surface) and the other
one (the STM or AFM tip) can also be approximately regarded as a large surface
but with a local structure at the contact. In the case of pulling the tip away
from the surface, a single apex atom connection or single atomic chain
connection may develop at the contacts, as has been shown experimentally
\cite{Xiao04267,Ohnishi,Yanson}. Similar to this situation, a recent experiment
\cite{au-chain} showed that by using directly STM atomic manipulation on
NiAl(110) surface one can assemble LML systems with precise contact atomic
structures, in which {\it single gold atomic chains} can be used as leads.

On the other hand, in future molecular electronics circuits, the interconnects
should be comparable in size to the functional devices and the best choice may
be some kind of one dimensional (1D) nanostructures. Recent experiments have
shown that 1D nanostructures like carbon nanotubes
\cite{Mceuen04272,Heer04281,Avouris04403} or semiconductor nanowires
\cite{Yang02553,Melosh03112,Beckman045921} are potentially ideal building
blocks for functional devices and interconnects in nanoelectronics. Their size
can be as small as $\sim$ 1 nm \cite{sinw1nm,cnt3a}.  From the different
experimental situations mentioned above, one can see the wide variety of
lead and contact structures and the significance of leads (interconnects) with
a nanometer diameter in future molecular electronics technologies.

With regard to theoretical modeling, a metal lead can be modeled by either an
infinite surface (for example, Refs.\ \ \onlinecite{infinite-lead-1,
infinite-lead-2, infinite-lead-3,diventra0,diventra1,diventra2}) or a very thin atomic wire (for example,
Refs.\ \ \onlinecite{nanowire-lead} and\ \ \onlinecite{ourwork1,ourwork2}). The
infinite-surface--lead
model is relevant to surface-STM/AFM experiments without the
pulling of the tip, while the thin atomic wire model is appropriate for the
possible nanowire or even atomic-chain interconnects in molecular electronics
circuits. On the other hand, the situation in MCB experiments or
surface-STM/AFM experiments, where a very thin nanowire connection or single
atomic chain connection is developed, is between these two
theoretical limits.

For a LML system with nanowire leads or connections around its
contacts, the behavior of the electron transport through the whole system will
not only depend on the molecule and contacts but also on the metallic lead
itself, in which quantum confinement effects will become significant because
of its small cross-section. In addition, for leads with a regular periodic
structure, there will be a strong quantum waveguide effect.  
This thin lead induced effect may be important for
understanding relevant experiments and for reasonable comparisons between
theoretical calculations and experiments. However, it has been unclear what the
detailed effect is and how significant it can be on the electron transport.
The main reason for this lack of understanding is that, to date, there have
been no calculations on electron transport through LML systems with leads of
different widths, from a thin wire to an infinite surface, in a consistent
manner, and therefore, there have been no theoretical analyses available concerning
quantum confinement effects in thin leads compared to 
infinite surface leads.

\begin{figure}[t]
\includegraphics[angle= 0,width=3.5cm]{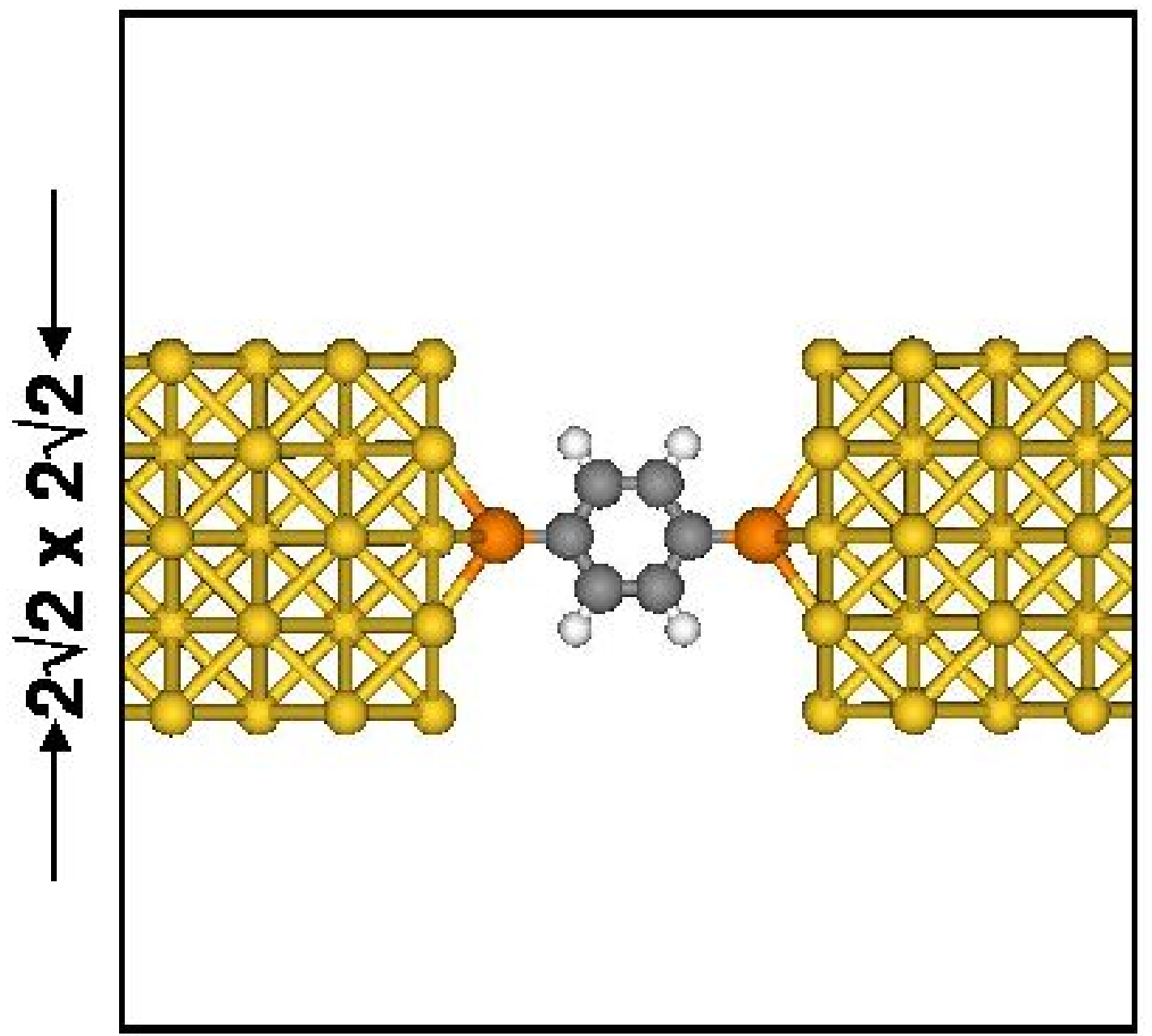} (a)
\includegraphics[angle= 0,width=3.5cm]{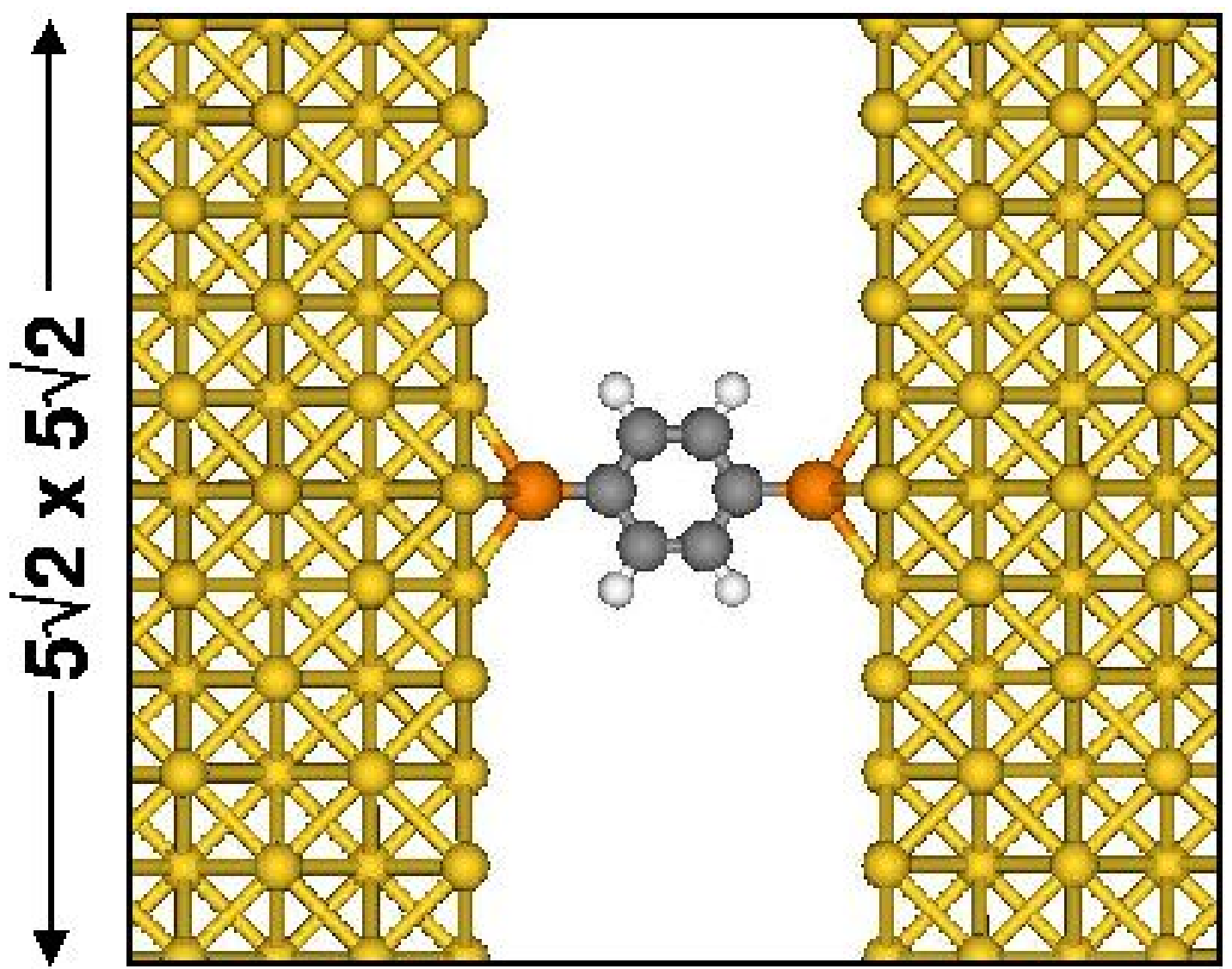} (b) \\
\includegraphics[angle= 0,width=7.5cm]{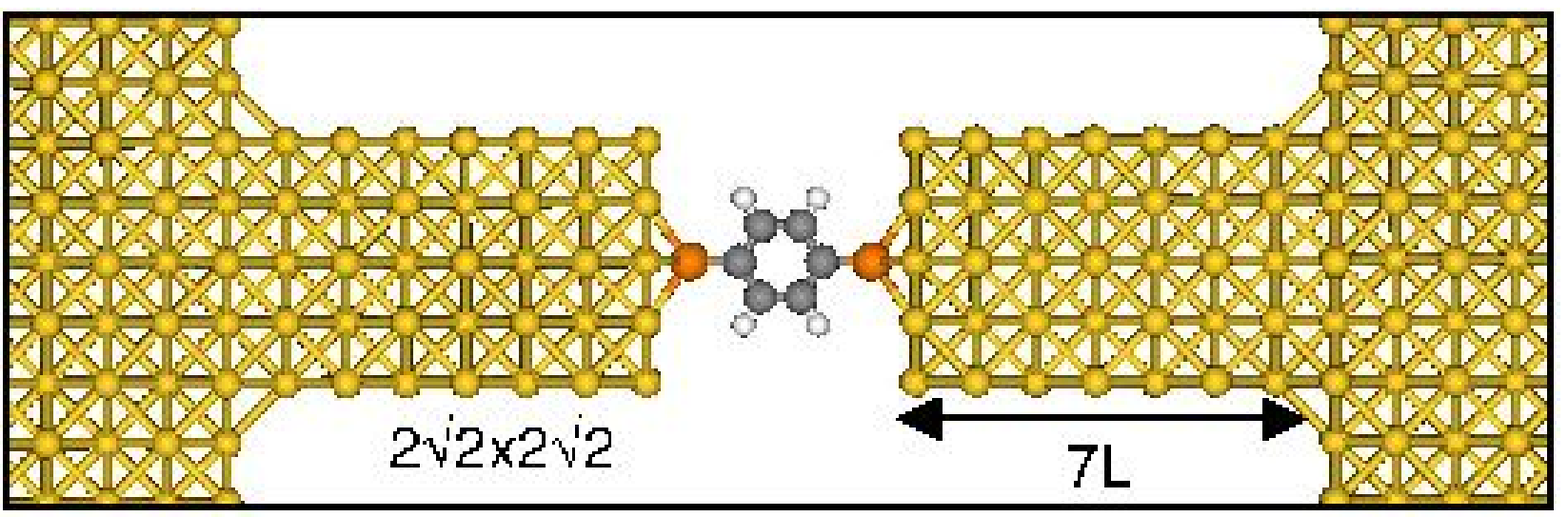} (c) \\
\includegraphics[angle= 0,width=3.5cm]{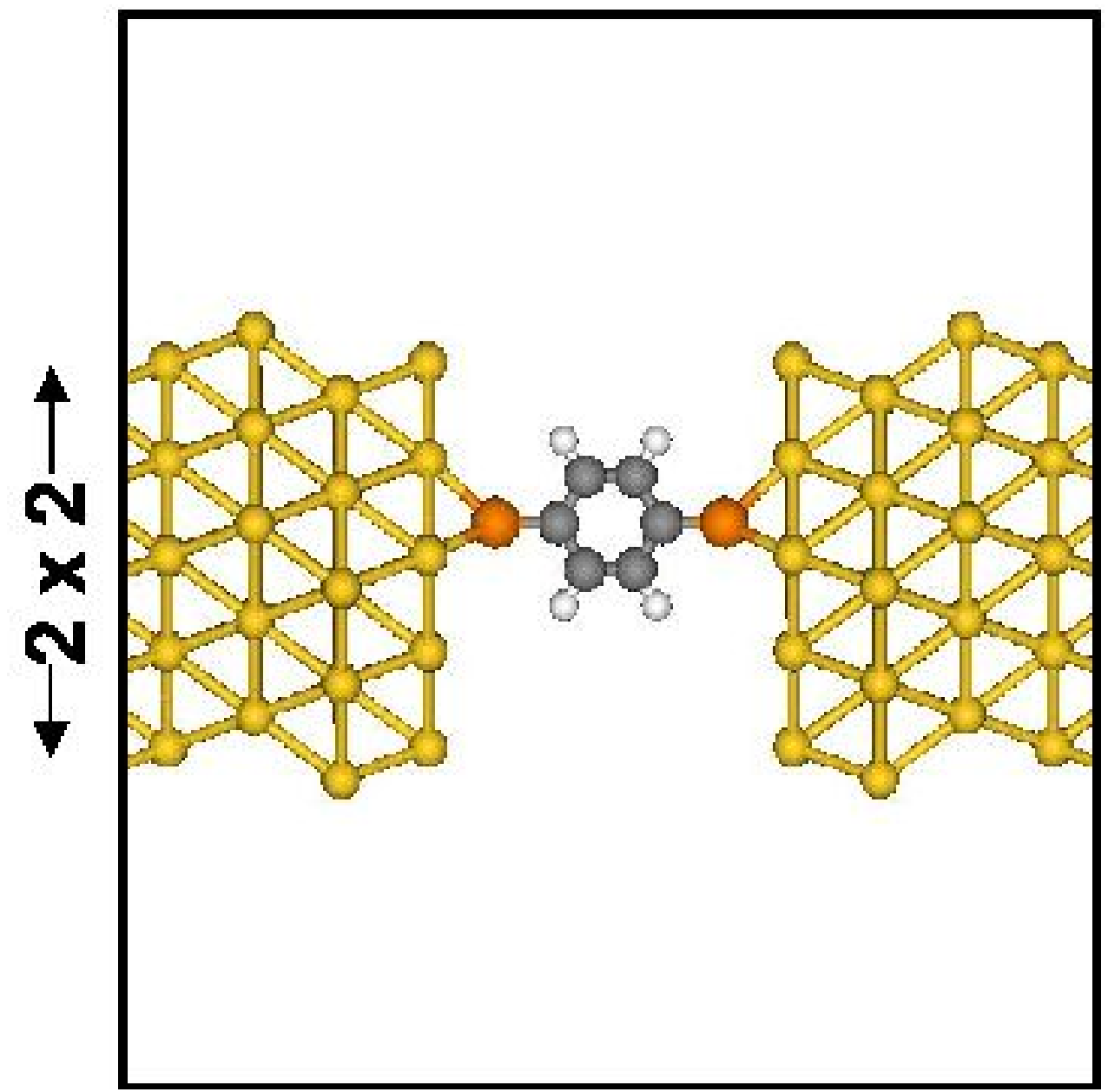} (d)
\includegraphics[angle= 0,width=3.5cm]{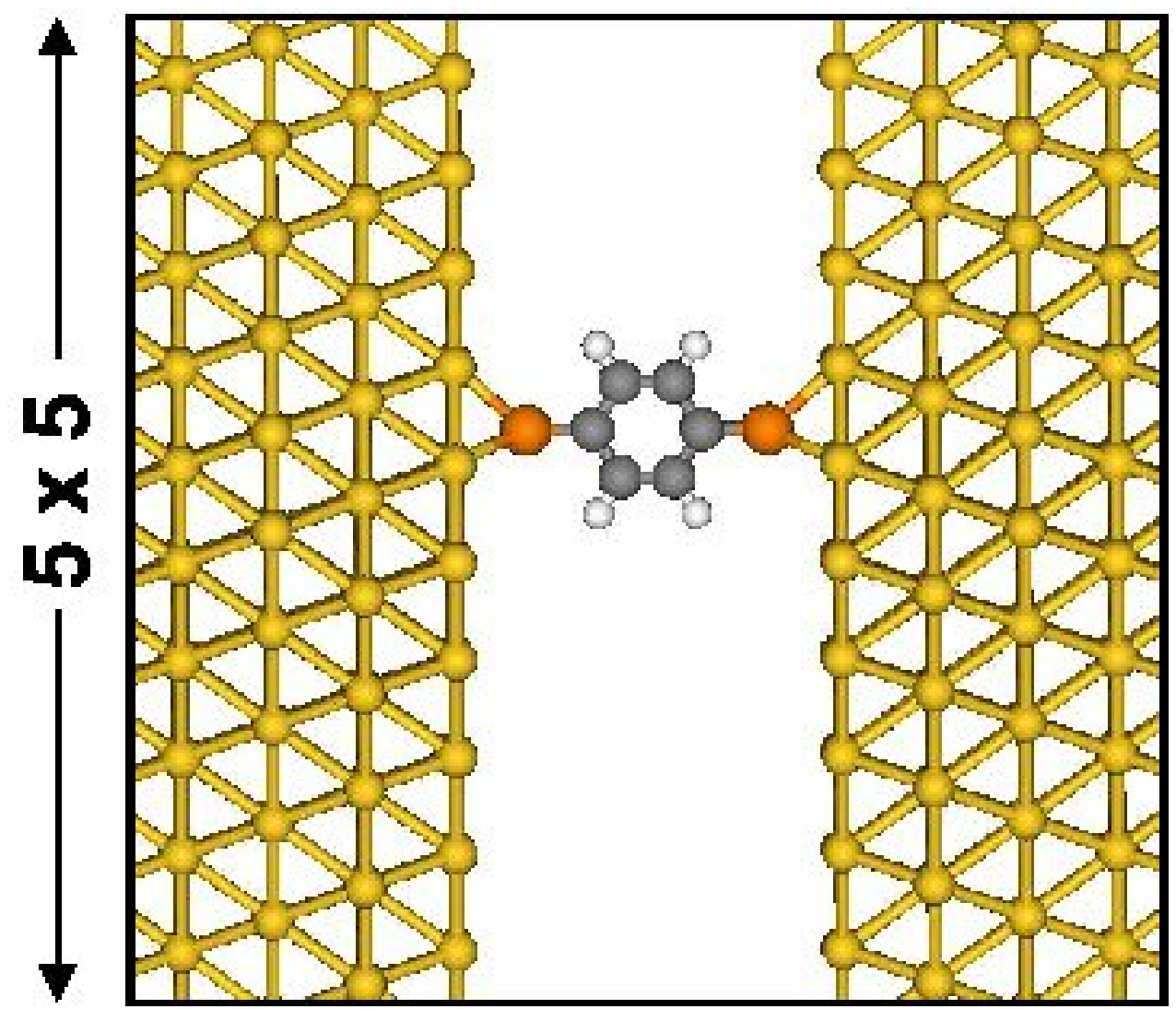} (e) \\
\includegraphics[angle= 0,width=7.5cm]{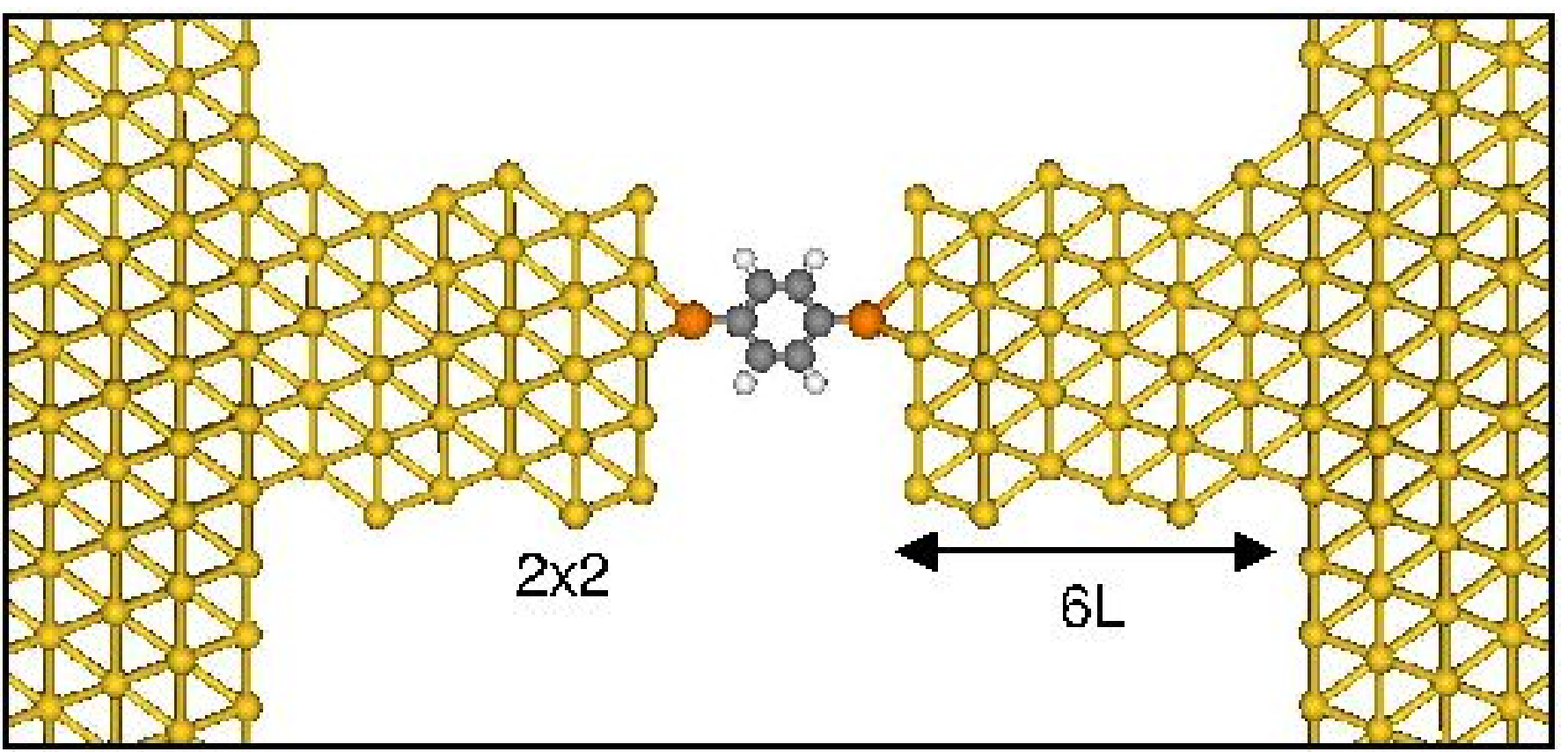} (f)
\caption{(color online) Examples of atomic structure of the device regions of the 
Au-benzenedithiol-Au systems in different models as explained in the text. 
The leads in (a)-(c) are in the (001)
orientation while those in (d)-(f) are in the (111) orientation. 
(a) and (d) are in Model A, where the finite cross-sections of the leads are
indicated. (b) and (e) are in Model B, where the frames
indicate the supercells for the infinite periodic surface leads. 
(c) and (f) are in Model C, where a nanowire is introduced
to connect the molecule to the extended lead; its width and length
are indicated: for instance, 7L in (c) means 7-atomic-layers long.}
\label{fig_str_lml}
\end{figure}

In this paper, we focus on exploring quantum confinement effects in leads
with a finite cross-section and on bridging the difference in contact atomic
structure between MCB or surface-STM/AFM experiments and theoretical
calculations. For these purposes, we carry out first-principles calculations of
molecular conductance of a Au-benzenedithiol-Au system adopting different atomic models for the leads and
contacts. Our strategy is to bring the two
theoretical limits together: We adopt a reliable model for the infinite surface
leads and use the result as a limit/reference to which the results from
systems with leads of different finite cross-sections are compared. Therefore,
we consider the following models for the lead: (1) (001) and (111) gold
nanowires with different widths and (2) infinite periodic Au(001) and Au(111)
surfaces. Furthermore, the nanowire structures that develop around the
lead-molecule contacts revealed in MCB experiments are simulated by introducing
a nanowire of varying length between the molecule and the infinite
periodic surface. Similarly, the single apex atom or single atomic chain
connections that occur in the surface-STM/AFM experiments are also simulated by
introducing a single Au atomic chain at each contact.

Our calculations show that a small cross-section lead causes large
oscillations in the electron transmission function [$T(E)$] because of
waveguide effects, and therefore this effect depends significantly on the lead
structure, such as different lead orientations. This oscillation slowly decays
along with the increase of the lead width, with the average approaching the limit given
by the corresponding infinite surface lead, for which the effect of different
lead orientations is significantly reduced. The local nanowire structure
around the contacts induces moderate fluctuations in $T(E)$,
keeping the main features unchanged if the nanowire is not too long. In contrast
to this, the Au atomic chain (including single apex Au atom)
connection at each contact leads to a more significant resonance in the equilibrium conductance.
Based on the above findings, together with some previous theoretical results, 
we discuss the relationship between {\it ab initio} theory and experiment in
molecular electronics. It is shown that for different contact structures
theoretical predictions of conductance are always much larger than relevant experimental
results, which has become an important issue in molecular electronics and 
needs to be addressed by further work.

\section{Modeling and computation}

\begin{figure}[t]
\includegraphics[angle= 0,width=8.0cm]{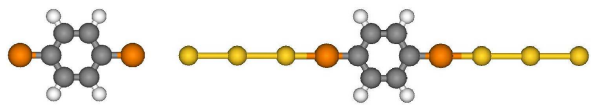}
\caption{(color online) Optimized atomic structure of the S-C$_6$H$_4$-S
molecule and the structure of Au$_3$-S-C$_6$H$_4$-S-Au$_3$ 
where two 3-Au-atom chains are attached (Au-Au distance is fixed at 2.65{\AA}).
The LML systems with the former have
been shown in Fig.~\ref{fig_str_lml}; those with the latter are similar but
have an 3-Au-atom chain at each contact (Model D).} \label{fig_str_mol2}
\end{figure}

We consider different models for the leads and lead-molecule contacts of a
Au-benzenedithiol-Au system: (A) leads with different finite cross-sections, 
(B) leads of infinite periodic surfaces, (C) periodic surface
leads with a nanowire connection of different lengths to the molecule, and 
(D) similar to (C) but with a single atomic chain
(including a single apex atom) at the contacts. Two lead
orientations, (001) and (111), are considered. Examples of atomic
structure of the device regions in the different models are shown in
Fig.~\ref{fig_str_lml}. Fig.~\ref{fig_str_lml} (a) shows the atomic structure
of a device region in Model A for the (001) orientation, which has 
leads with a finite cross-section, 2$\sqrt{2} \!\times\! 2\sqrt{2}$.
Fig.~\ref{fig_str_lml} (b) is a (001) system in Model B, which has leads of  
5$\sqrt{2} \!\times\! 5\sqrt{2}$ periodic surface.
Fig.~\ref{fig_str_lml} (c) shows the device region of a (001) system 
in Model C, where a 7-atomic-layer 2$\sqrt{2} \!\times\!
2\sqrt{2}$ nanowire is introduced to connect the molecule to the 4$\sqrt{2}
\!\times\! 4\sqrt{2}$ periodic surface lead, to simulate the possible
experimental situations in MCB experiments mentioned previously.
Figs.~\ref{fig_str_lml} (d) -- (f) show the atomic structures of the
counterparts for the (111) lead orientation. For all the structures the Au-Au bond
length in the leads is fixed at 2.89 {\AA} (i.e., experimental bulk bond
length) and the molecules are adsorbed at
the hollow site of the Au(001) or Au(111) surface with the molecule-surface
separation optimized. As can be seen in
Fig.~\ref{fig_str_lml}, we adopt very large supercells for Model B. Because of
the large separation between the molecule and its images (larger than 12{\AA}),
the interference among supercells will be very small; therefore, this model is a good approximation to a lead consisting of an infinitely large surface, as we
will demonstrate later. The single atomic chain connection in
Model D is equivalent to changing the molecule from S-C$_6$H$_4$-S to
Au$_n$-S-C$_6$H$_4$-S-Au$_n$, as shown in Fig.~\ref{fig_str_mol2}, where a
3-Au-atom-chain ($n$=3) is attached on each side. We consider $n$= 1, 2, and 3
in our calculation for Model D.

We adopt a nonequilibrium Green function (NEGF) \cite{negf1,negf2} method combined with density
functional theory (DFT) \cite{dft2} electronic structure calculations to investigate the electron transport \cite{datta1,datta2,pal1,mcdcal,transiesta,trank1}. Specifically,\cite{trank1} we divide an
infinite LML system into three parts: left lead, right lead, and
device region ($C$) which contains the molecule and large parts of the left and
right leads (devices regions shown in Fig.~\ref{fig_str_lml}),
so that the molecule-lead interactions (couplings) can be fully accommodated.
Unlike some other theoretical calculations adopting infinite surface leads
\cite{infinite-lead-1,infinite-lead-2,infinite-lead-3,diventra0,diventra1,diventra2} in which the device
region and leads are treated on different theoretical levels, here all the
subsystems are treated on exactly the same footing. For a
steady state situation in which region $C$ is under a bias $V_b$ (zero or
finite), its density matrix ($\mathbf{D}_C$) and Hamiltonian ($\mathbf{H}_C$)
can be determined self-consistently by DFT+NEGF techniques
\cite{datta1,datta2,pal1,mcdcal,transiesta,trank1}. The Kohn-Sham
wave-functions are used to construct a single-particle Green function from
which the transmission coefficient at any energy is calculated. The
conductance, $G$, then follows from a Landauer-type relation. The detailed
computational techniques have been described previously \cite{trank1}.

For the electronic structure calculation, we use DFT and adopt a numerical
basis set to expand the wave functions \cite{siesta}. A single zeta plus
polarization basis set (SZP) is adopted for all atomic species. Our test 
calculation for a small system shows that the result of the SZP calculation
has only minor differences from that of a calculation using a higher level double
zeta plus polarization basis set (DZP). We make use of
optimized Troullier-Martins pseudopotentials \cite{tmpp} for the atomic cores.
The PBE version of the generalized gradient approximation (GGA) \cite{pbe} is
used for the electron exchange and correlation. The atomic structure of the
isolated molecule and the molecule-lead separation are fully optimized using
the higher level DZP basis set.

\section{Leads: finite and infinite cross-section}

\begin{figure}[t]
\includegraphics[angle=-90,width=8.0cm]{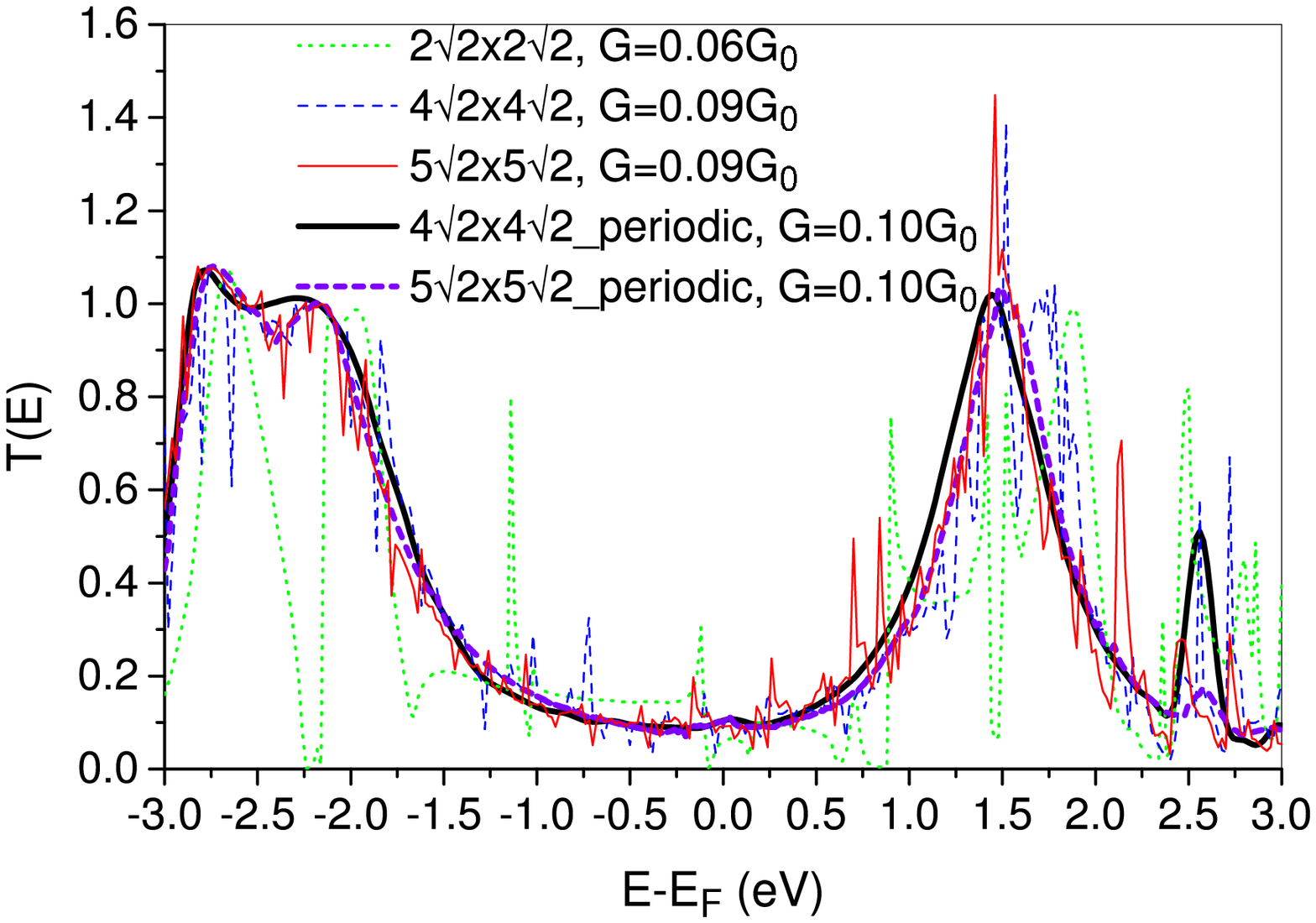} (a)
\includegraphics[angle=-90,width=8.0cm]{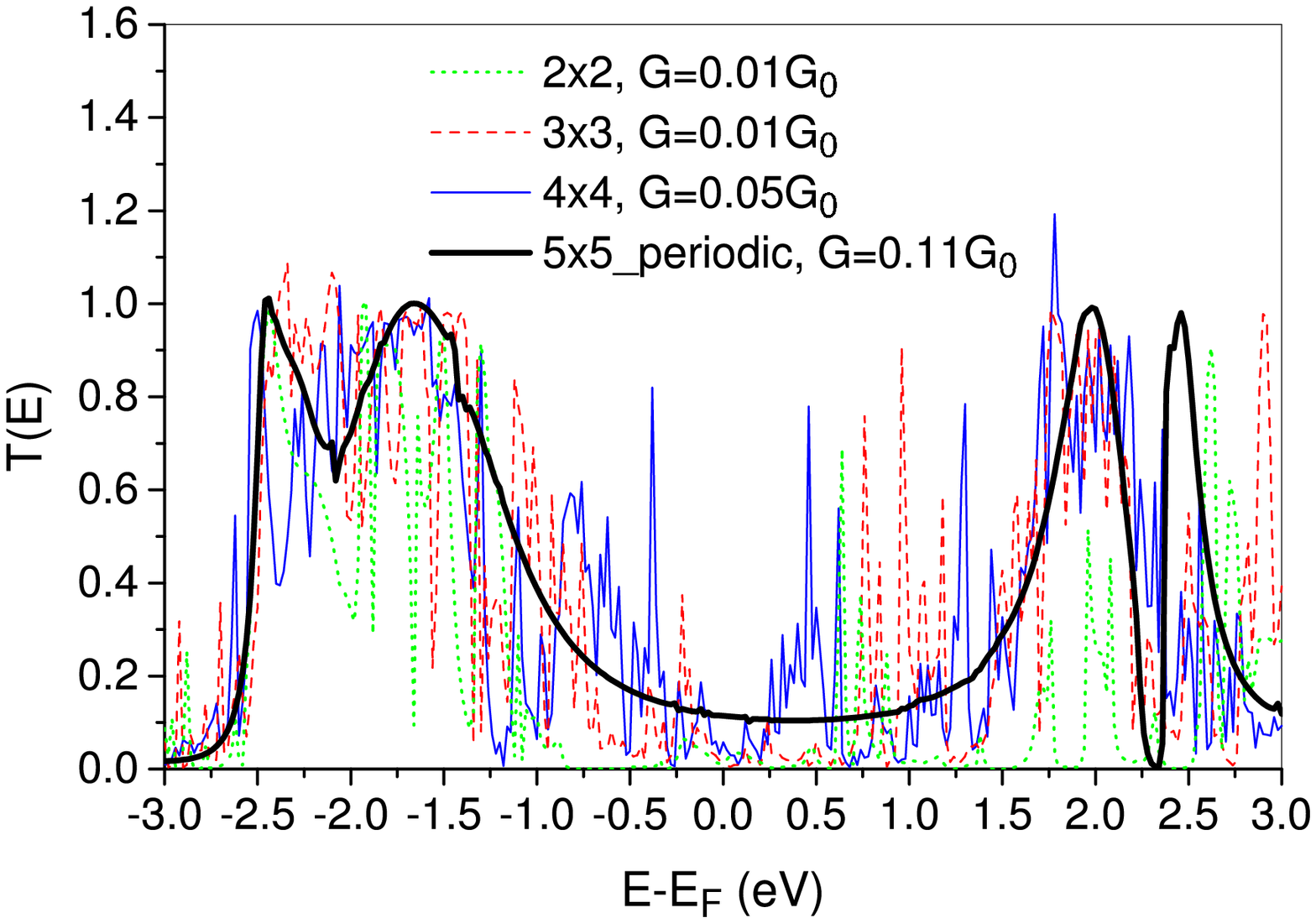} (b)
\caption{(color online) Calculated $T(E)$ functions for the different LML systems 
in Models A and B [examples of their atomic structures are shown 
in Figs.~\ref{fig_str_lml} (a), (b) for (001) orientation, and (d), (e)
for (111)]. The width of the leads with a finite
cross-section and the size of the supercells used for the periodic surface
leads are indicated in the legends. Also indicated is the equilibrium
conductance $G$. } \label{fig_t_0au}
\end{figure}

As has been mentioned previously, the use of nanowire leads with small cross
sections in molecular electronics or nanoelectronics is very promising.
Therefore, it is important to investigate its possible consequences. So far it
has been unclear what the effect of a very thin lead is and how significant it
can be on the electron transport through the LML system. Here we investigate
this issue by calculating the electron transmission adopting Model A with
different lead widths in comparison with Model B. 

In Figs.~\ref{fig_t_0au} (a)
and (b) we show $T(E)$ for Models A and B for the (001) and (111)
lead orientations, respectively.
In Model B we use a periodic surface lead to simulate an infinitely large
surface lead. Obviously, for this purpose we have the issue of convergence of
the in-plane size of the supercell. If the size is small the molecule will
interact with its images and the result will vary remarkably with the
increasing size. Here we adopt a very large in-plane size [($4\sqrt{2} \!\times\!
4\sqrt{2}$) for the (001) orientation and $5 \!\times\! 5$ for the (111)], as a
result, the direct inter-molecule interactions have been completely removed
because of the large inter-molecule separation ( $>$ 12 {\AA} ) and the use of
the basis functions with a finite range. In order to check the convergence we
compare the result of $T(E)$ from the $4\sqrt{2} \!\times\! 4\sqrt{2}$ surface lead
to that from an even larger $5\sqrt{2} \!\times\! 5\sqrt{2}$ surface lead. >From the
results in Fig.~\ref{fig_t_0au} (a) we see that the two curves are
very close to each other, indicating that the in-plane size of these periodic
surface leads is already large enough to simulate an infinitely large surface
lead.

Fig.~\ref{fig_t_0au} (a) shows that the $T(E)$ function of Model A depends
strongly on its lead width. The small cross-section, $2\sqrt{2} \!\times\!
2\sqrt{2}$, leads to large sharp oscillations in the $T(E)$ function. Along
with the increase of the lead width to $4\sqrt{2} \!\times\! 4\sqrt{2}$ and
$5\sqrt{2} \!\times\! 5\sqrt{2}$, the amplitude of this oscillation decays and a large-scale
structure appears, whose average is approaching the limit given by the periodic surface
leads. In spite of the overall large difference in $T(E)$ between Model B and
Model A with thin leads, the difference in their equilibrium conductance is
quite small for the (001) orientation.

The results for the (111) lead orientation are summarized in
Fig.~\ref{fig_t_0au} (b). Here the oscillations in $T(E)$
for Model A are even much larger than those in the (001) case and the
convergence with the lead width is much slower: even the result given by the
lead with $4 \!\times\! 4$ cross-section does not show the large-scale structure
given by the $5 \!\times\! 5$ periodic surface lead. Related to this much stronger
oscillation, here the equilibrium conductance given by Model A is significantly
smaller (by about one order of magnitude) than that given by Model B, in
contrast to the (001) case. An interesting thing to notice is that although in
Model A the $T(E)$ functions for the (001) and (111) orientations are very
different, they become similar in Model B, as shown by the solid lines in
Fig.~\ref{fig_t_0au}. The only major difference is in the relative position of
the Fermi energy in the gap, which is a result of the different charge
transfer around the (001) and (111) contacts.

\begin{figure}[t]
\includegraphics[angle=  0,width=8.0cm]{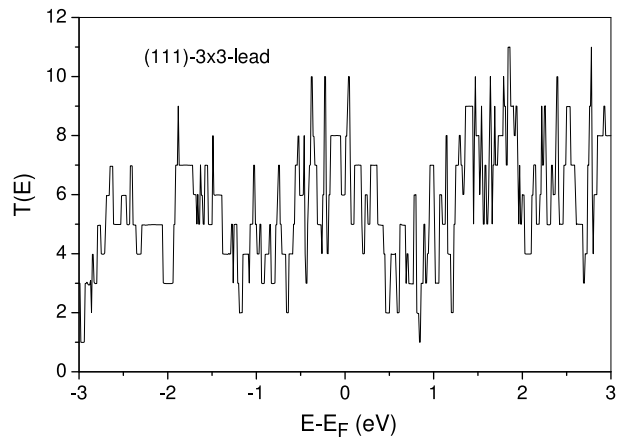}
\caption{Transmission function of the pure infinitely long (111)-3x3 lead
(nanowire). Note the step-function structure because of waveguide effects.}
\label{fig_t_111_lead}
\end{figure}

\begin{figure}[t]
\includegraphics[angle=-90,width=8.0cm]{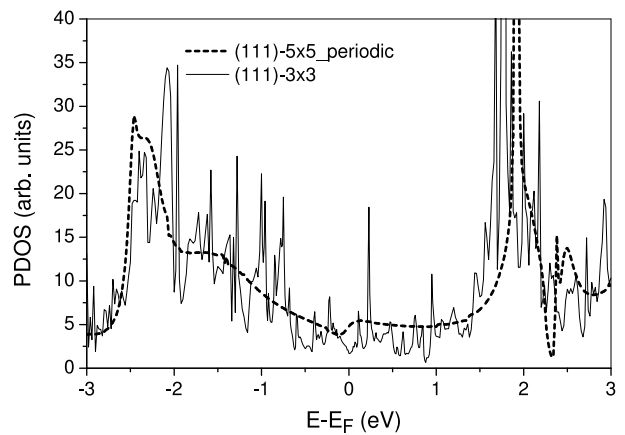}
\caption{Projected density of states on the S-C$_6$H$_4$-S molecule for the
(111) LML systems with the 3$\!\times\!$3 finite-cross-section lead and the
5$\!\times\!$5 periodic surface lead. Note the large oscillation in the former.}
\label{fig_pdos_111}
\end{figure}

The large oscillations in $T(E)$ function given by Model A for both the (001)
and (111) lead orientations can be understood by considering the waveguide
effect of the leads. In Model A the leads are infinitely long periodic
nanowires. As a result of the transversal quantum confinement and the periodic
structure, the electron transmission coefficient through these leads is always
an integer. However, because of their complicated
atomic structure the transverse modes in these 1D waveguides will have a
complicated dependence on the electron energy, resulting in sharp step-function
oscillations in their $T(E)$ functions. To show this clearly, we calculate
$T(E)$ for the pure infinitely long (111) lead with $3 \!\times\! $3 cross
section. The result in Fig.~\ref{fig_t_111_lead} clearly shows the expected large
step-function oscillation. After scattering by the molecule,
this strongly oscillating transverse mode spectra will cause large oscillations
in the projected density of states (PDOS) on the molecule, as shown in
Fig.~\ref{fig_pdos_111}. Obviously, this oscillating PDOS combined with the
molecule-lead coupling will give a strongly oscillating $T(E)$ spectra for the
whole LML system, as we already see in Fig.~\ref{fig_t_0au} (b). The behavior
of a nanowire waveguide will depend critically on its cross-section atomic
structure. Because the (111) nanowire leads are more irregular in atomic
structure and have much lower symmetry than the (001) nanowire leads, the
oscillations in the $T(E)$ of the (111) LML systems will be
stronger and the convergence with respect to the lead width will be slower
compared to the (001) case, as can be seen by comparing Figs.~\ref{fig_t_0au}
(a) and (b).

As the lead gets wider, the number of the transverse modes in the lead increases, and
in addition, the number of modes that are coupled to the molecule
also increases. At the same time, the average transmission through the whole
system remains the same. Thus the transmission from a given mode in the
lead through the molecule will go down. Hence the threshold
singularity associated with that mode will also decrease. 
Therefore, as the lead gets wider and wider, the oscillation structure
in the full T(E) should become both smaller and finer. 
However, this expected decay is not so clear when we increase the width of the (001) lead 
from $4\sqrt{2} \!\times\! 4\sqrt{2}$ to $5\sqrt{2} \!\times\! 5\sqrt{2}$ and that of the
(111) lead from $4 \!\times\! 4$ to $5 \!\times\! 5$ (see Fig.~\ref{fig_t_0au}).
This indicates that (1) this kind of decay is slow and the cross-section of the leads used in our
calculations is still too small to clearly show it, and therefore (2) this quantum waveguide effect
may be measurable for real nanometer-sized leads/interconnects in molecular
electronics.

\begin{figure}[t]
\includegraphics[angle=-90,width=7.0cm]{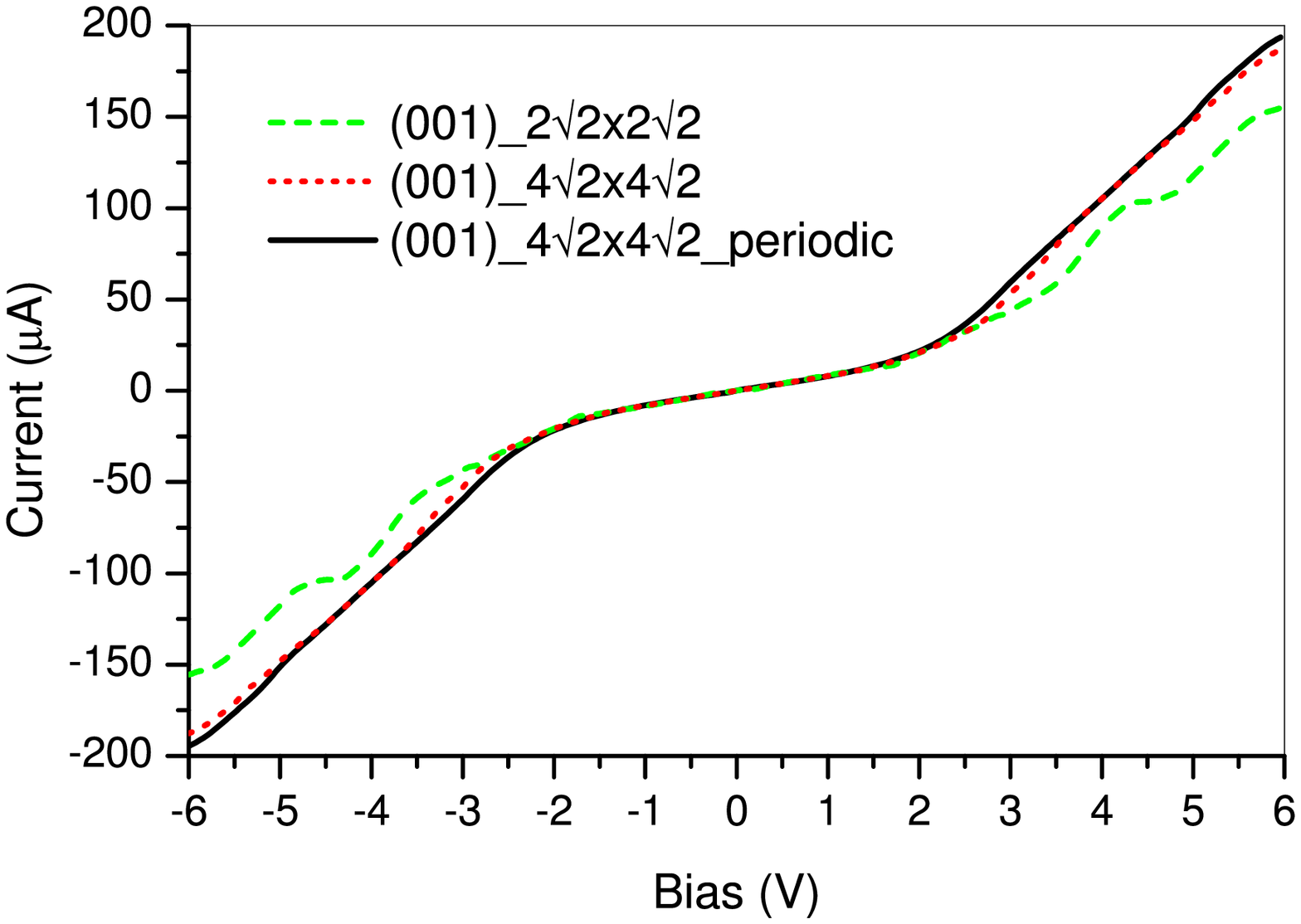} (a)
\includegraphics[angle=-90,width=7.0cm]{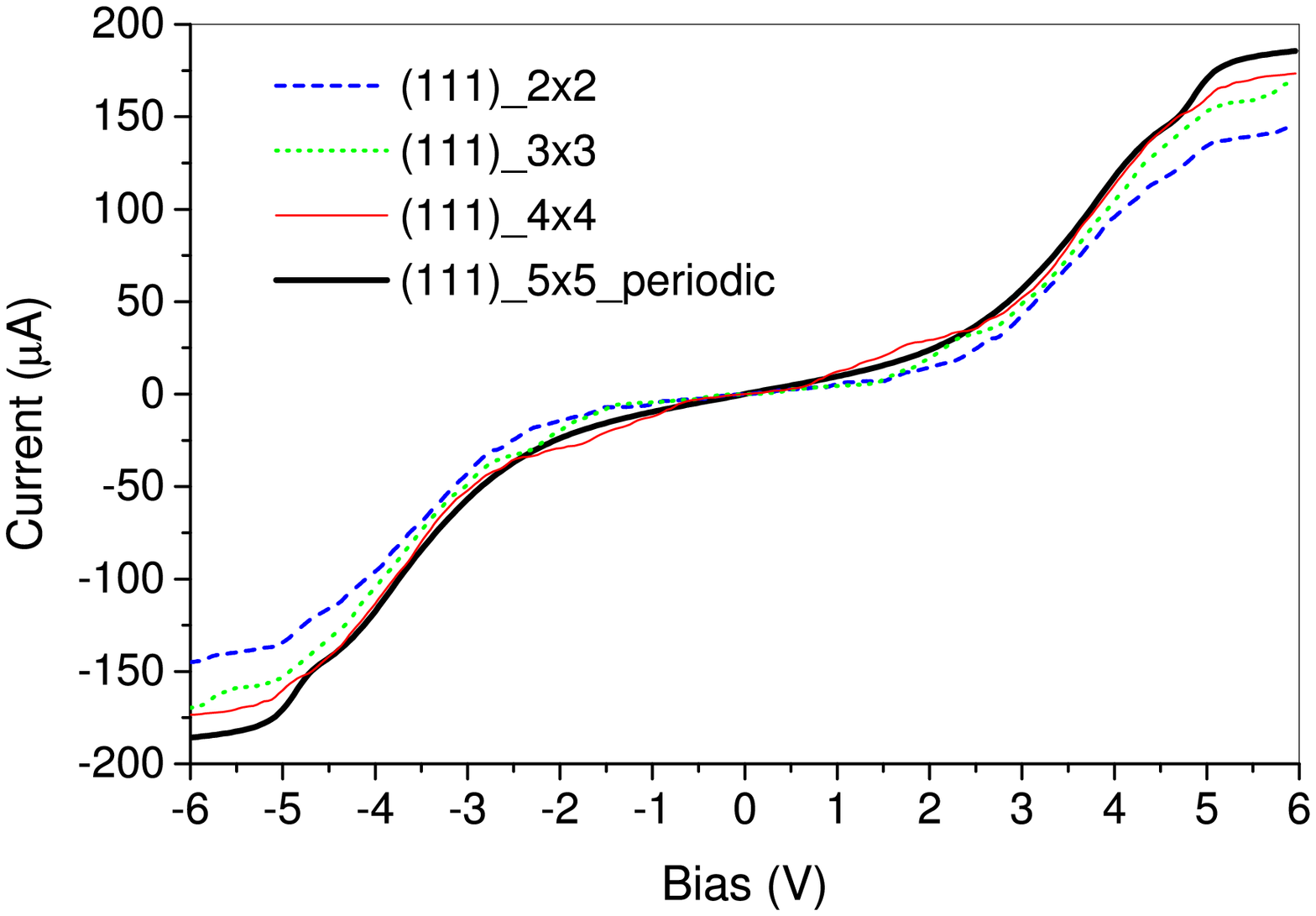} (b)
\caption{(color online) $I$-$V$ curves of the different LML systems with (a) (001) leads and (b)
(111) leads [see Fig.~\ref{fig_str_lml} (a), (b) and (d), (e)], which are calculated from
the $T(E)$ functions in Fig.~\ref{fig_t_0au} neglecting bias-induced effects in
$T(E)$. The cross-sections of the leads are indicated in the legends.
} \label{fig_iv_0au}
\end{figure}

The large oscillations in $T(E)$ of Model A may also have some effects
on its $I$-$V$ characteristics. To show the possible effects while avoiding the
too large computational cost due to the different large lead widths, we
calculate the $I$-$V$ curves directly from the $T(E)$ functions in
Fig.~\ref{fig_t_0au}, i.e., bias-induced changes in $T(E)$ are
neglected. The results given in Fig.~\ref{fig_iv_0au} show that despite the
significant dependence of $T(E)$ on the structure and cross-section of the leads in Model A, the $I$-$V$ characteristic is not so sensitive
to these factors: even for very thin nanowire leads the main feature of the
$I$-$V$ characteristic has already been captured.

\section{Nanowire connection around the contacts}

\begin{figure}[t]
\includegraphics[angle=-90,width=8.0cm]{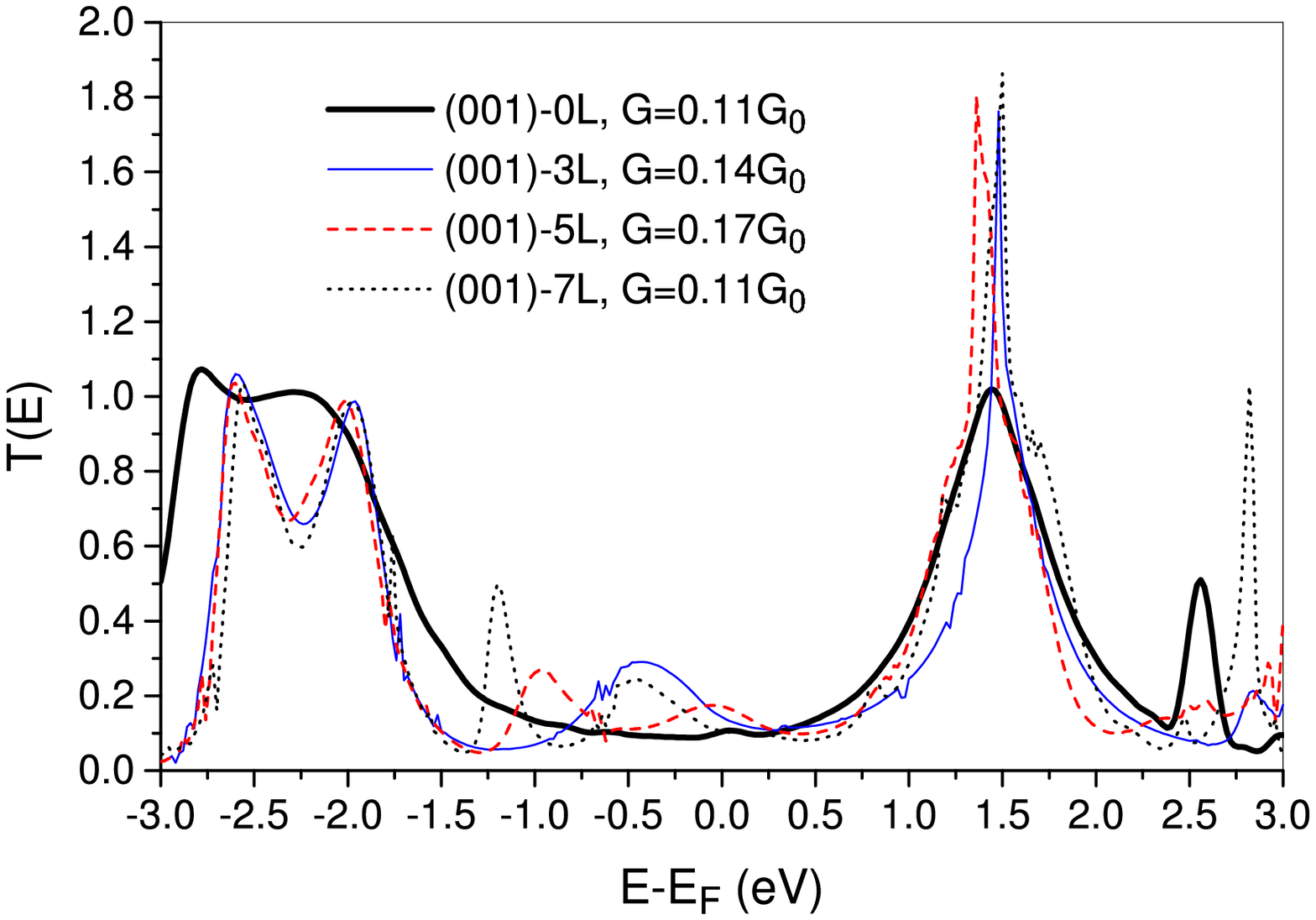} (a)
\includegraphics[angle=-90,width=8.0cm]{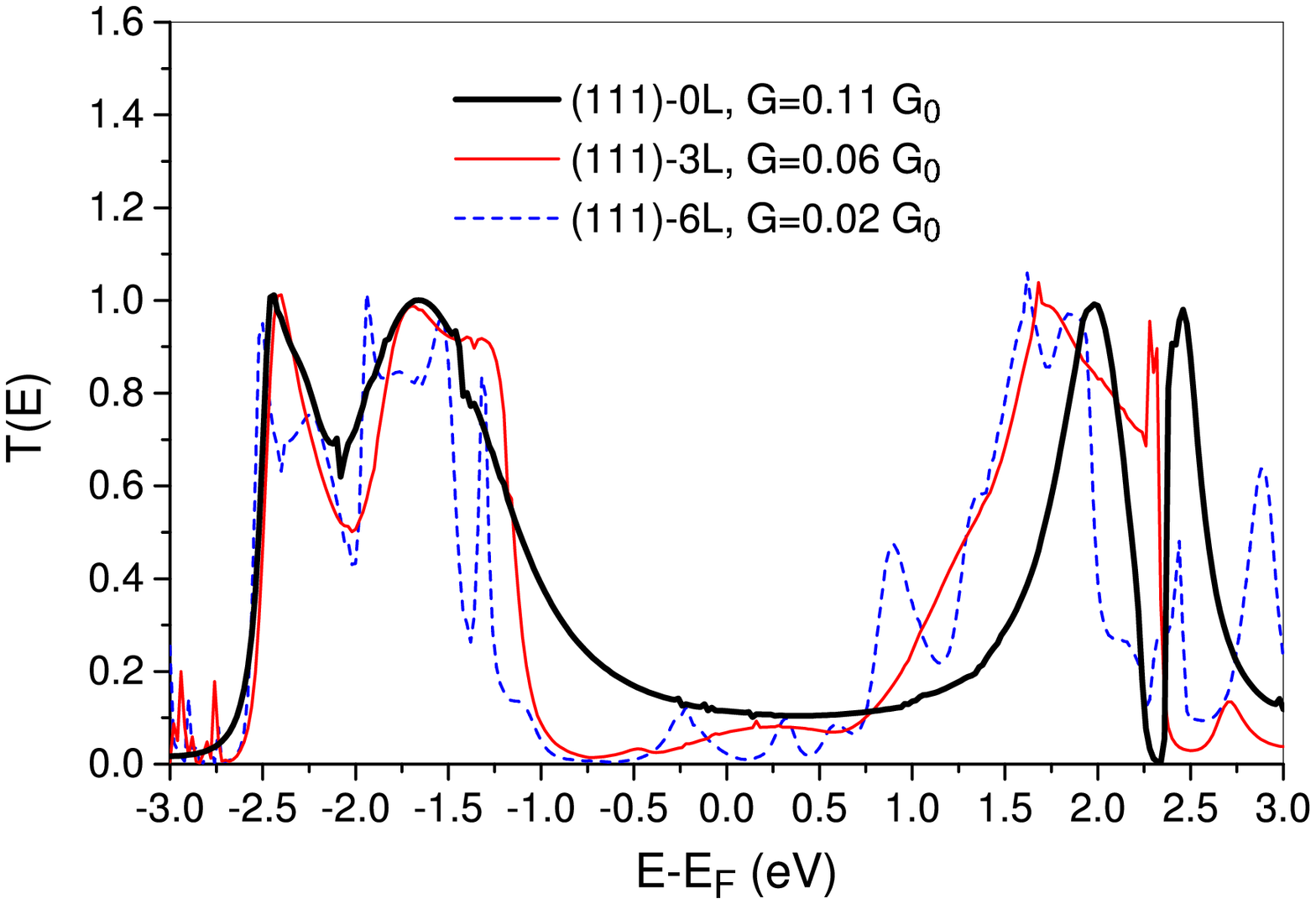} (b)
\caption{(color online) Transmission functions of the LML systems with a nanowire connecting
the molecule to the extended lead, as shown in Figs.~\ref{fig_str_lml} (c) and
(f). The length of the nanowire and the equilibrium conductance of different
LML systems are listed in the legends. } \label{fig_t_xL_0au}
\end{figure}

While having been adopted extensively in
previous theoretical calculations, leads with a tiny cross-section
\cite{nanowire-lead,ourwork1,ourwork2,trank1} and with an infinite cross-section
\cite{infinite-lead-1,
infinite-lead-2,infinite-lead-3,diventra0,diventra1,diventra2} are the two theoretical
limits, which correspond to the Model A and Model B, respectively, in this
paper. The real situation in MCB experiments is probably between these two,
where a finite-length nanowire is usually developed connecting the molecule to
the extended electrode of a LML system. To simulate this situation, here we
consider a nanowire with different lengths, which connects the benzene molecule
to an infinite periodic surface (i.e., Model C). The cross-section of the
nanowire is set to be $2\sqrt{2} \!\times\! 2\sqrt{2}$ for the (001) system and $2
\!\times\! 2$ for the (111) system. The atomic structures of these systems are
shown in Fig.~\ref{fig_str_lml} (c) and (f), where the length of the nanowire
is denoted by the number of atomic layers it contains (i.e., 7L means a
nanowire consisting of 7 atomic layers). 

The calculated $T(E)$ functions are
shown in Fig.~\ref{fig_t_xL_0au}. As can be seen, the introduction of the
nanowire connection around the contacts causes noticeable changes in
transmission coefficient over the whole energy range. This is because the
finite-length nanowire depresses some transverse modes and enhances some
others. In spite of this, the main feature of the $T(E)$ functions still
remains for the lengths of the nanowires studied here. Keeping the behavior of
Model A in mind, one can image that, along with the increase of the length of
the nanowire, the induced fluctuation in $T(E)$ will become larger and larger.
This is just the case as we can see in Fig.~\ref{fig_t_xL_0au}, especially for
the (111) system. Also similar to Model A, here the (111) nanowire induces more
fluctuation than the (001) nanowire does, especially in the equilibrium
conductance as listed in the legends of Fig.~\ref{fig_t_xL_0au}.

\section{Effects of atomic chain connections around the contacts}

\begin{figure}[t]
\includegraphics[angle=-90,width=7.5cm]{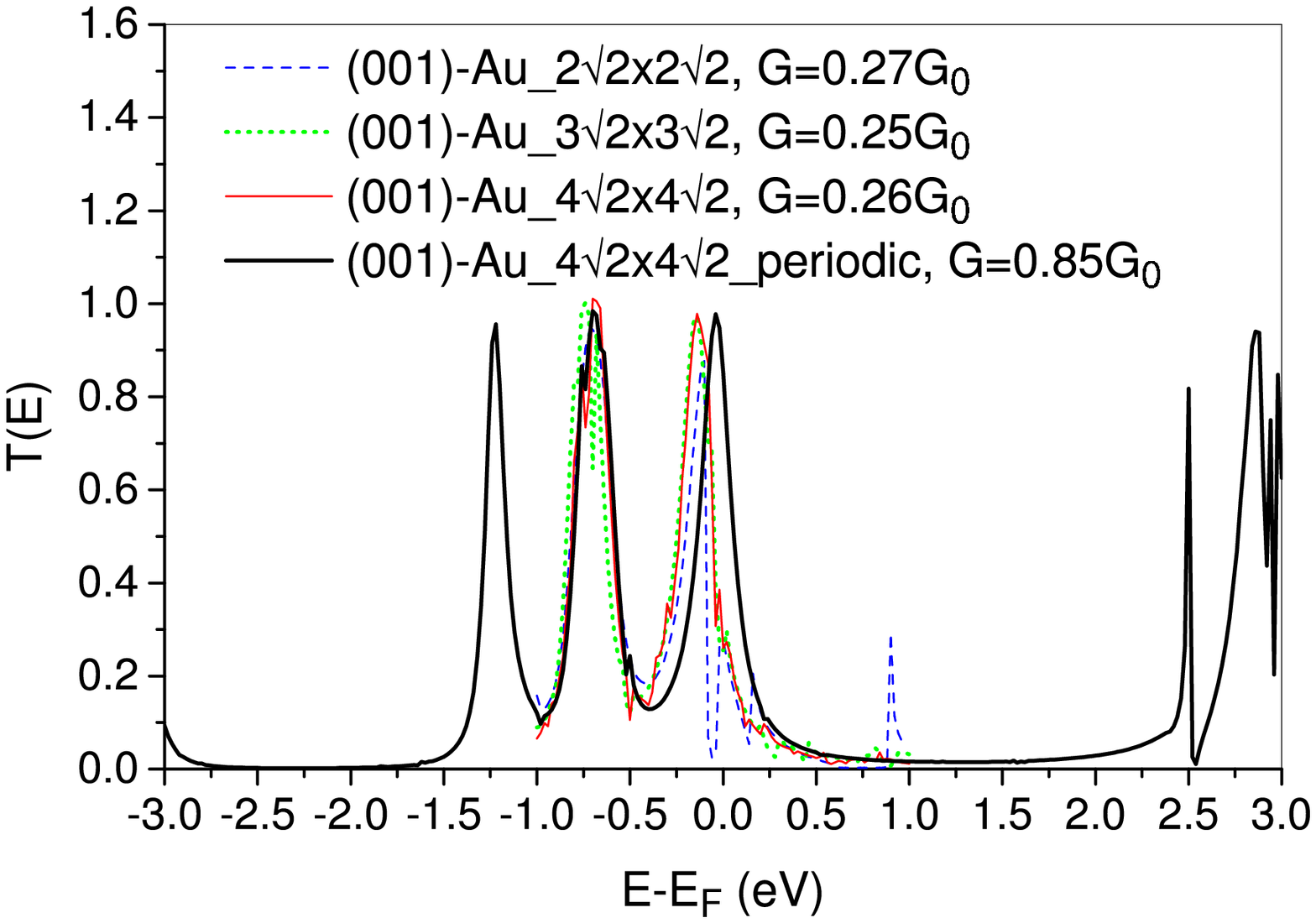} (a)
\includegraphics[angle=-90,width=7.5cm]{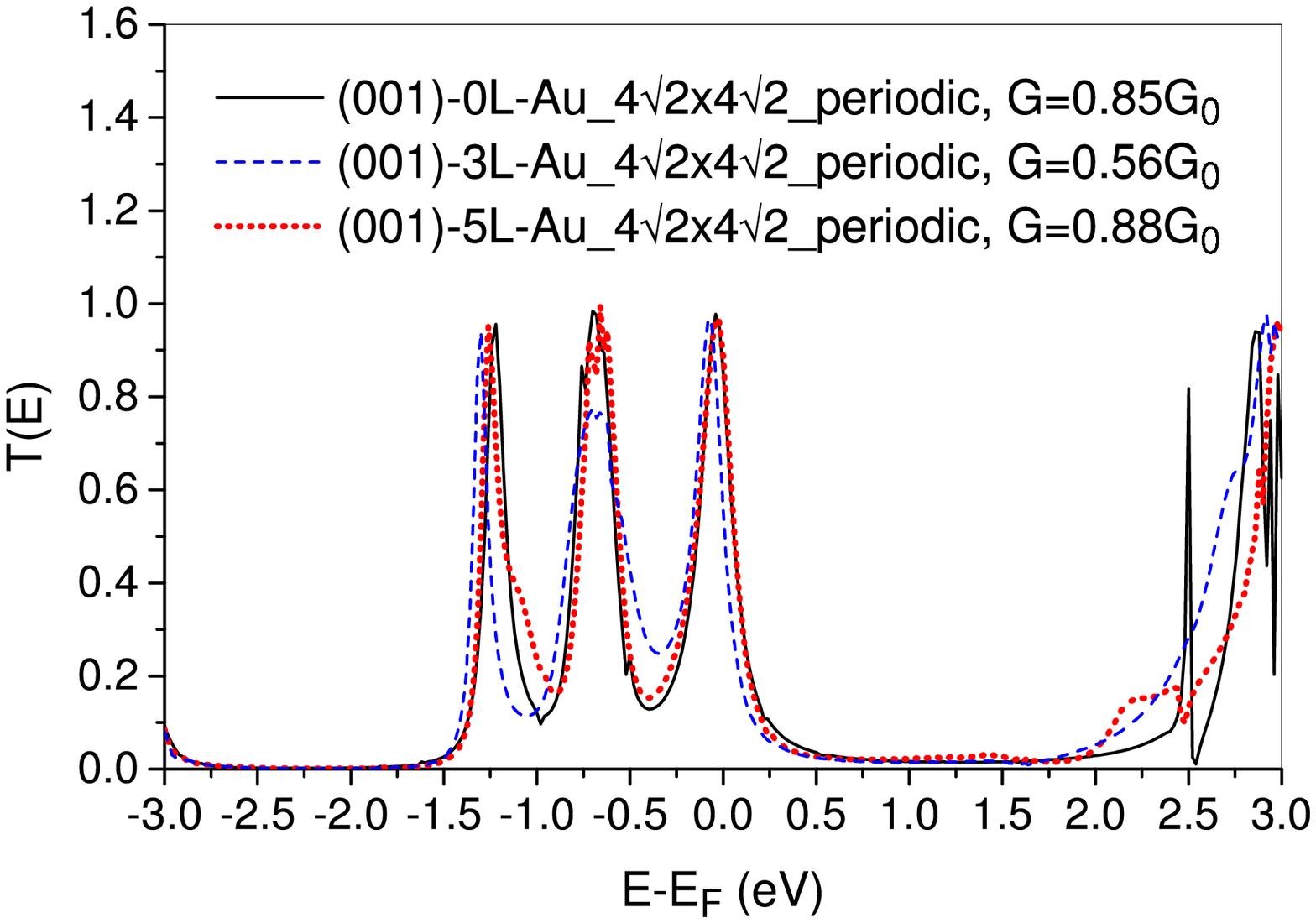} (b)
\includegraphics[angle=-90,width=7.5cm]{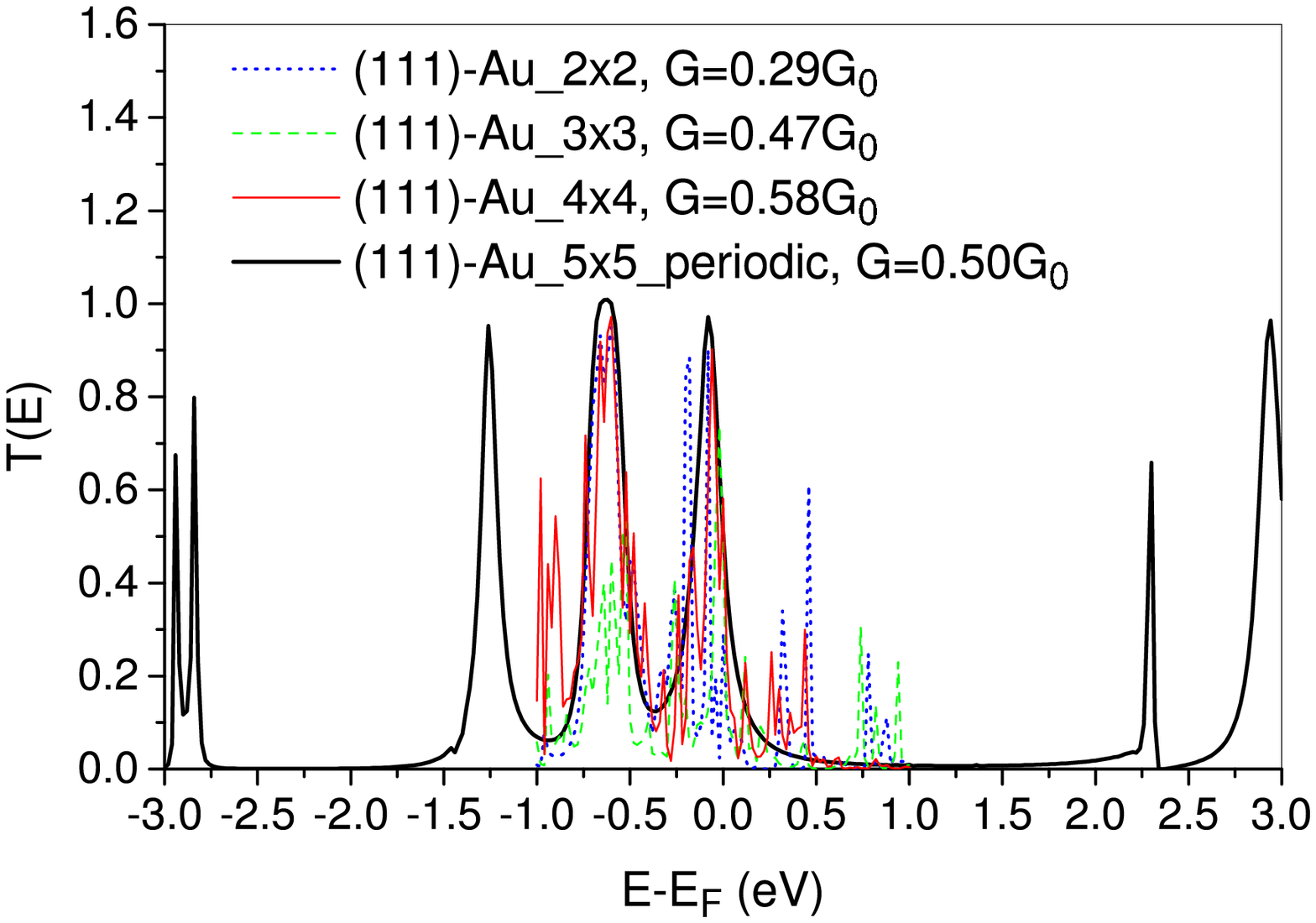} (c)
\caption{(color online) Transmission functions of the LML systems shown in
Figs.~\ref{fig_str_lml} (a)-(e), but with an apex Au atom at each
contact, as indicated in Fig.~\ref{fig_str_mol2}. The structures of these LML
systems and their equilibrium conductances are indicated in the legends. Note
the large resonance peak around the Fermi energy and the similar $T(E)$
structures for all the different systems due to the introduction of the
additional Au atom at each contact.} \label{fig_t_1au}
\end{figure}

Besides the large scale changes in atomic structure that we have discussed
above, there are other possibly more localized changes around contacts in MCB
experiments. One is atomic fluctuation (roughness) of the break surfaces, and
as a result, a molecule may be connected through an apex atom rather directly
to the flat surface. In surface-STM/AFM experiments, if we pull the tip away
from the surface a single apex connection or a single atomic chain connection
will develop \cite{Xiao04267,Ohnishi,Yanson}. 
Here we simulate this situation by adding a Au atomic chain,
Au$_n$ ($n$= 1, 2, and 3), at each contact between the molecule and the
infinite surface lead. 
%

\begin{figure}[t]
\includegraphics[angle=-90,width=8.0cm]{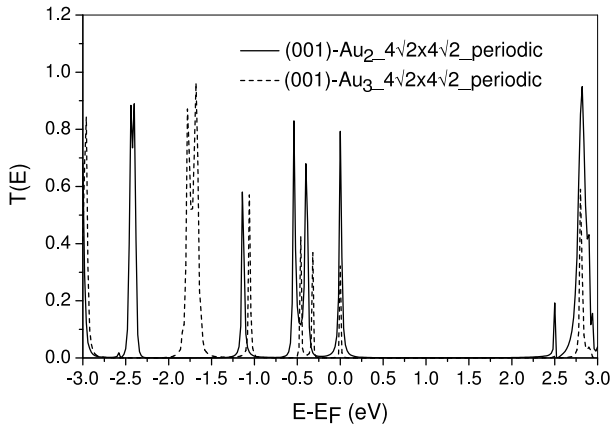}
\caption{Transmission functions of the LML system shown in
Figs.~\ref{fig_str_lml} (b), but with an atomic chain of two (solid line) and
three (dashed line) Au atoms at each contact, as indicated in
Fig.~\ref{fig_str_mol2} for the 3-Au-atom chain. Note that the resonance peak
around the Fermi energy remains as the number of Au atoms in the
chain changes.} \label{fig_t_23au}
\end{figure}


First, let us look at the case of $n \!=\! 1$ which means an additional apex Au
atom at each contact. For the systems with small leads (Model A) its
effect was previously investigated \cite{ourwork1,ourwork2}. It was found that presence
of the additional Au atoms increases significantly the equilibrium
conductance due to a resonance caused by a molecular level aligned with the Au Fermi energy. Here we further
investigate the effect for Models B and C, and to check if the conclusion is
affected by the strong oscillating behavior of Model A.

In Fig.~\ref{fig_t_1au} we show $T(E)$ functions for
Models A, B, and C. The introduction of
the additional Au atom changes totally the $T(E)$ functions and leads to a
large resonance peak around the Fermi energy for all the different models,
independent of the different structures in each model. Thus, the conclusion about the effect
of the additional Au atom, which was reached previously for Model A with small
lead widths \cite{ourwork1,ourwork2}, is a general result regardless of the lead and
contact structures.

For Model A in the (001) orientation, Fig.~\ref{fig_t_1au} (a) shows that the
additional Au atom stabilizes the $T(E)$ structure even for the smallest lead
width, and the result of Model A converges quickly to that of Model B. For Model
A in the (111) orientation, this effect also exists but is weaker, as shown
in Figs.~\ref{fig_t_1au} (c), because of the much stronger oscillating
behavior. For all the different structures in Models B and C, including the
different lead orientations and the different lengths of connecting nanowire
in Model C, the introduction of the additional Au atom changes
totally $T(E)$ and leads to very similar results. This indicates
that in this system it is the local change in contact structure that is
important in determining the transport properties of the whole LML system.

The results for $n$=2 and 3 are given in Fig.\ \ref{fig_t_23au}. The increase in
the length of the Au atomic chain narrows the peaks in $T(E)$, keeping
the main feature unchanged except for a new peak around -2.5 eV for $n$=2 or -1.7
eV for $n$=3. The large resonance peak around the Fermi energy remains although
its height is reduced as $n$ increases.

These results are particularly interesting in light of the experiments
and conclusions of Ref. \cite{Xiao04267}. There measurements of the
conductance through single molecules of benzenedithiol and
benzenedimethanethiol by using a STM technique were reported.  The conductance found was much larger
than previous experimental results \cite{bj1}.  The paper goes on to
state that the new results are in agreement with theoretical
expectations, citing Ref.\:\onlinecite{diventra1}. The claimed agreement
with theory is quite significant because the big discrepancy in
conductance between {\it ab initio} theory
\cite{diventra0,diventra1,diventra2,taylor03,infinite-lead-1,ourwork1}
and experiments
\cite{bj1,zhou97,reichert03,kubatkin03,rawlett02,kushmerick03} (about
two orders of magnitude) is a long standing issue in the field of
molecular electronics.

The experiment of Ref.\:\onlinecite{Xiao04267} differs from previous ones in
that additional gold atoms may be present at the molecule-gold contact.
The paper states that the distance over which a molecular junction can
be stretched for both benzenedithiol and
benzenedimethanethiol is between 0.3 and 0.6 nm. This
distance is not how long the molecules themselves are stretched but
rather is due to reconfiguration of the gold contacts: When a gold atom
at the molecule-gold contact is pulled out of the electrode, a nearby
surface gold atom moves behind the first atom. Further pulling can cause
a third atom to move behind the second one and form a linear atomic
chain. Based on this observation and analysis, the authors suggest that
the molecule is connected to the electrode through an apex Au atom. 
Indeed, 0.3 nm is in very good agreement with the distance change due to
one apex Au atom at each of the two contacts.

The theoretical configuration for comparison should certainly take into
account the molecule-metal contact region, and, in particular, must
include an apex Au atom. The work used for comparison \cite{diventra1},
however, did not include an apex Au atom and used the jellium model for
the electrodes which does not give a good account of the molecule-metal
contact region.

Comparing the experimental results of Ref.\:\onlinecite{Xiao04267} with the
results of this Section, we find that the experimental value is still
much smaller than the theoretical results, by one to two orders of
magnitude. Therefore the long standing discrepancy between experiment
and theory remains, and urgently needs to be addressed by further work.

\section{Summary}

In an effort to explore the effect of a small lead cross-section and to
bridge the difference in atomic structure between experiment and theory, we
have investigated different models for the lead and contact structures of a
Au-benzenedithiol-Au system: leads with different finite cross-sections, leads consisting of
infinite surfaces, and surface leads with a local nanowire or a  
single atomic chain connecting the molecule.
The findings are as follows:

(1) Waveguide effects in leads with a small cross-section will lead to large sharp oscillations in the
$T(E)$ function, which depend significantly on the lead structure (orientation). These oscillations slowly get smaller as the lead width increases, with the average approaching the limit given by the infinite
surface leads, for which the effect of different lead structures is
significantly reduced. The effect of this strong oscillation on $I$-$V$
characteristics is, however, relatively weak.

(2) The local nanowire structure around the contacts will induce
noticeable fluctuations in $T(E)$ function. However, the main feature of the
$T(E)$ function will remain if the nanowire structure is not too long, as those
considered in this paper.

(3) In contrast to the above effect, the single atomic chain
(including a single apex atom) at
each contact leads to a large robust resonance in the conductance.

We appreciate valuable conversations with Rui Liu. This work was supported in part by the NSF (DMR-0103003).

\end{document}